\documentclass{llncs}
\usepackage[T1]{fontenc} 
\usepackage{lmodern}
\usepackage{graphicx}
\usepackage{cite}
\usepackage{amsmath}
\usepackage{amsfonts}
 
\usepackage{color}
\usepackage{multicol}
\usepackage{multirow}
\usepackage{enumerate}
\usepackage{bm}

\newcommand{\R}{\mathbb{R}}
\newtheorem{heuristic}{Heuristic Definition}

\begin{document}

\title{Problem formulation for truth-table invariant cylindrical algebraic decomposition by incremental triangular decomposition}
\author{Matthew England$^1$ \and Russell Bradford$^1$ \and Changbo Chen$^2$ \and James H. Davenport$^1$ \and Marc Moreno Maza$^3$ \and David Wilson$^1$}
\institute{
$^1$ University of Bath, Bath, BA2 7AY, U.K. \\
$^2$ Chongqing Key Laboratory of Automated Reasoning and Cognition,
Chongqing Institute of Green and Intelligent Technology, 
CAS, Chongqing, 400714, China. \\
$^3$ University of Western Ontario, London, N6A 5B7, Canada. \\ \vspace*{0.1in}
\email{ \texttt{ \{ R.J.Bradford, J.H.Davenport, M.England, D.J.Wilson \}@bath.ac.uk}, \\ \texttt{changbo.chen@hotmail.com}, \texttt{moreno@csd.uwo.ca} }
}
\maketitle

\begin{abstract}
Cylindrical algebraic decompositions (CADs) are a key tool for solving problems in real algebraic geometry and beyond.  We recently presented a new CAD algorithm combining two advances: truth-table invariance, making the CAD invariant with respect to the truth of logical formulae rather than the signs of polynomials; and CAD construction by regular chains technology, where first a complex decomposition is constructed by refining a tree incrementally by constraint.  We here consider how best to formulate problems for input to this algorithm.  We focus on a choice (not relevant for other CAD algorithms) about the order in which constraints are presented.   We develop new heuristics to help make this choice and thus allow the best use of the algorithm in practice.  We also consider other choices of problem formulation for CAD, as discussed in CICM 2013, revisiting these in the context of the new algorithm. 


\end{abstract}


\section{Introduction} 
\label{SEC:Intro}

A \emph{cylindrical algebraic decomposition} (CAD) is: a \emph{decomposition} of $\R^n$, meaning a collection of cells which do not intersect and whose union is $\R^n$; \emph{cylindrical}, meaning the projections of any pair of cells with respect to a given variable ordering are either equal or disjoint; and, \emph{(semi)-algebraic}, meaning each cell can be described using a finite sequence of polynomial relations.

CAD was introduced by Collins in \cite{Collins1975}, such that a given set of polynomials had constant sign on each cell.  This meant that a single sample point for each cell was sufficient to conclude behaviour on the whole cell and thus it offered a constructible solution to the problem of quantifier elimination.  Since then a range of other applications have been found for CAD  including robot motion planning \cite{SS83II}, epidemic modelling \cite{BENW06}, parametric optimisation \cite{FPM05}, theorem proving \cite{Paulson2012} and reasoning with multi-valued functions and their branch cuts \cite{DBEW12}.

In \cite{BCDEMW14} the present authors presented a new CAD algorithm combining two recent advances in CAD theory: construction by first building a cylindrical decomposition of complex space, incrementally refining a tree by constraint \cite{CM12b}; and the idea of producing CADs such that given formulae has invariant truth on each cell \cite{BDEMW13}.  Experimental results in \cite{BCDEMW14} showed this new algorithm to be superior to its individual components and competitive with the state of the art.
We now investigate the choices that need to be made when using the new algorithm.  

We conclude the introduction with the necessary background theory and then in Section \ref{SEC:ECOrd} we demonstrate how constraint ordering affects the behaviour of the algorithm.  No existing heuristics discriminate between these orderings and so we develop new ones, which we evaluate in Section \ref{SEC:Results}.  In Section \ref{SEC:Other} we consider other issues of problem formulation, revisiting \cite{BDEW13} in the context of the new algorithm.   

\subsection{Background on CAD}
\label{SUBSEC:Background}

The first CAD algorithm, introduced by Collins \cite{Collins1975} with a full description in \cite{ACM84I}, works in two phases.  First in the \textit{projection} phase a projection operator is repeatedly applied to the set of  polynomials (starting with those in the input), each time producing another set in one fewer variables.  Then in the \textit{lifting} phase CADs are built incrementally by dimension.  First $\mathbb{R}^1$ is decomposed according to the real roots of the univariate polynomials.  Then $\mathbb{R}^2$ is decomposed by repeating the process over each cell in $\mathbb{R}^1$ using the bivariate polynomials evaluated at a sample point, and so on.
%
%
%
Collins' original projection operator was chosen so that the CADs produced could be concluded \emph{sign-invariant} with respect to the input polynomials, meaning the sign of each polynomial on each cell is constant.
%

Such decompositions can contain far more information than required for most applications, which motivated CAD algorithms which consider not just polynomials but their origin.  For example, when using CAD for quantifier elimination partial CAD \cite{CH91} will avoid lifting over a cell if the solution there is already apparent.  Another key adaptation is to make use of an \emph{equational constraint} (EC): an equation logically implied by an input formula.  The algorithm in \cite{McCallum1999} ensures sign-invariance for the polynomial defining an EC, any any other polynomials only when that constraint is satisfied.
A discussion of the first 20 years of CAD research is given in \cite{Collins1998}.  Some of the subsequent developments are discussed next, with others including the use of certified numerics when lifting \cite{Strzebonski2006, IYAY09}.

\subsection{TTICAD by regular chains}
\label{SUBSEC:RCTTICAD}

In \cite{BCDEMW14} we presented a new CAD algorithm, referred to from now on as \texttt{RC-TTICAD}.  It combined the following two recent advances.
\begin{description}
\item[Truth-table invariant CAD:]  A TTICAD is a CAD produced relative to a list of formulae such that each has constant truth value on every cell.  
\end{description}
The first TTICAD algorithm was given in \cite{BDEMW13}, where a new projection operator was introduced which acted on a set of formulae, each with an EC.  

\newpage

TTICADs are useful for applications involving multiple formulae like branch cut analysis (see for example Section 4 of \cite{EBDW13}), but also for building truth-invariant CADs for a single formula if it can be broken into sub-formulae with ECs.  The algorithm was extended in \cite{BDEMW14} so that not all formulae needed ECs, with savings still achieved if at least one did.  These algorithms were implemented in the freely available \textsc{Maple} package \textsc{ProjectionCAD} \cite{England13b}.  
\begin{description}

\item[CAD by regular chains technology:] A CAD may be built by first forming a \emph{complex cylindrical decomposition} (CCD) of $\mathbb{C}^n$ using triangular decomposition by regular chains, which is refined to a CAD of $\mathbb{R}^n$.  

\end{description}
This idea to break from projection and lifting was first proposed in \cite{CMXY09}.  In \cite{CM12b} the approach was improved by building the CCD incrementally by constraint, allowing for competition with the best projection and lifting implementations.  Both algorithms are implemented in the \textsc{Maple}  \textsc{RegularChains} Library, with the algorithm from \cite{CMXY09} currently the default CAD distributed with \textsc{Maple}.

\texttt{RC-TTICAD} combined these advances by adapting the regular chains computational approach to produce truth-table invariant CCDs and hence CADs.  This new algorithm is specified in \cite{BCDEMW14} where experimental results showed a \textsc{Maple} implementation in the \textsc{RegularChains} Library as superior to the two advances independently, and competitive with the state of the art.  The CCD is built using a tree structure which is incrementally refined by constraint.  ECs are dealt with first, with branches refined for other constraints in a formula only is the ECs are satisfied.  
Further, when there are multiple ECs in a formula branches can be removed when the constraints are not both satisfied.  See \cite{BCDEMW14, CM12b} for full details.

The incremental building of the CCD offers an important choice on problem formulation: in what order to present the constraints?  Throughout we use $A \rightarrow B$ to mean that $A$ is processed before $B$, where $A$ and $B$ are polynomials or constraints defined by them.  Existing CAD algorithms and heuristics do not discriminate between constraint orderings \cite{DSS04, BDEW13} and so a new heuristic is required to help make an intelligent choice.

\section{Constraint ordering}
\label{SEC:ECOrd}

The theory behind \texttt{RC-TTICAD} allows for the constraints to be processed in any order.  However, the algorithm as specified in \cite{BCDEMW14} states that \textbf{equational constraints should be processed first}.  This is logical as we need only consider the behaviour of non-ECs when corresponding ECs are satisfied, allowing for savings in computation.

We also advise \textbf{processing all equational constraints from a formula in turn}, i.e. not processing one, then moving to a different formula before returning to another in the first.  Although not formally part of the algorithm specification, this should avoid unnecessary computation by identifying when ECs have a mutual solution before more branches have been created.

There remain two questions to answer with regards to constraint ordering:\\
\textbf{Q1)} In what order to process the formulae?\\
\textbf{Q2)} In what order to process the equational constraints within each formula?

\subsection{Illustrative example}
\label{SUBSEC:Example}

\begin{figure}[t]
\vspace*{-15pt}
\caption{Visualisations of the four TTICADs which can be built using \texttt{RC-TTICAD} for Example \ref{ex:Learning}.  The figures on the top have $\phi_1 \rightarrow \phi_2$ and those on the bottom $\phi_2 \rightarrow \phi_1$.  The figures on the left have $f_1 \rightarrow f_2$ and those on the right $f_2 \rightarrow f_1$.
}
\label{fig:Learning}
\centering
\includegraphics[width=0.48\textwidth]{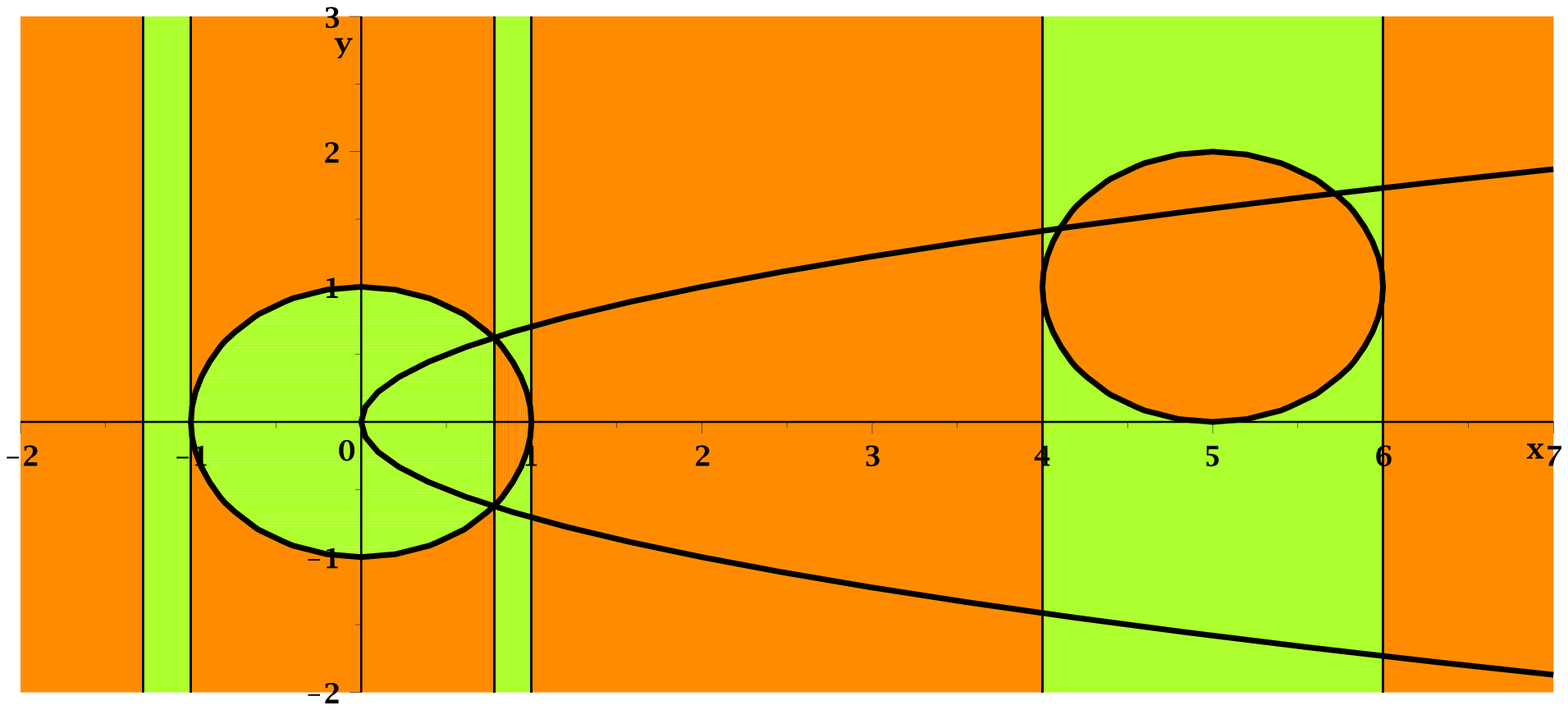}
\includegraphics[width=0.48\textwidth]{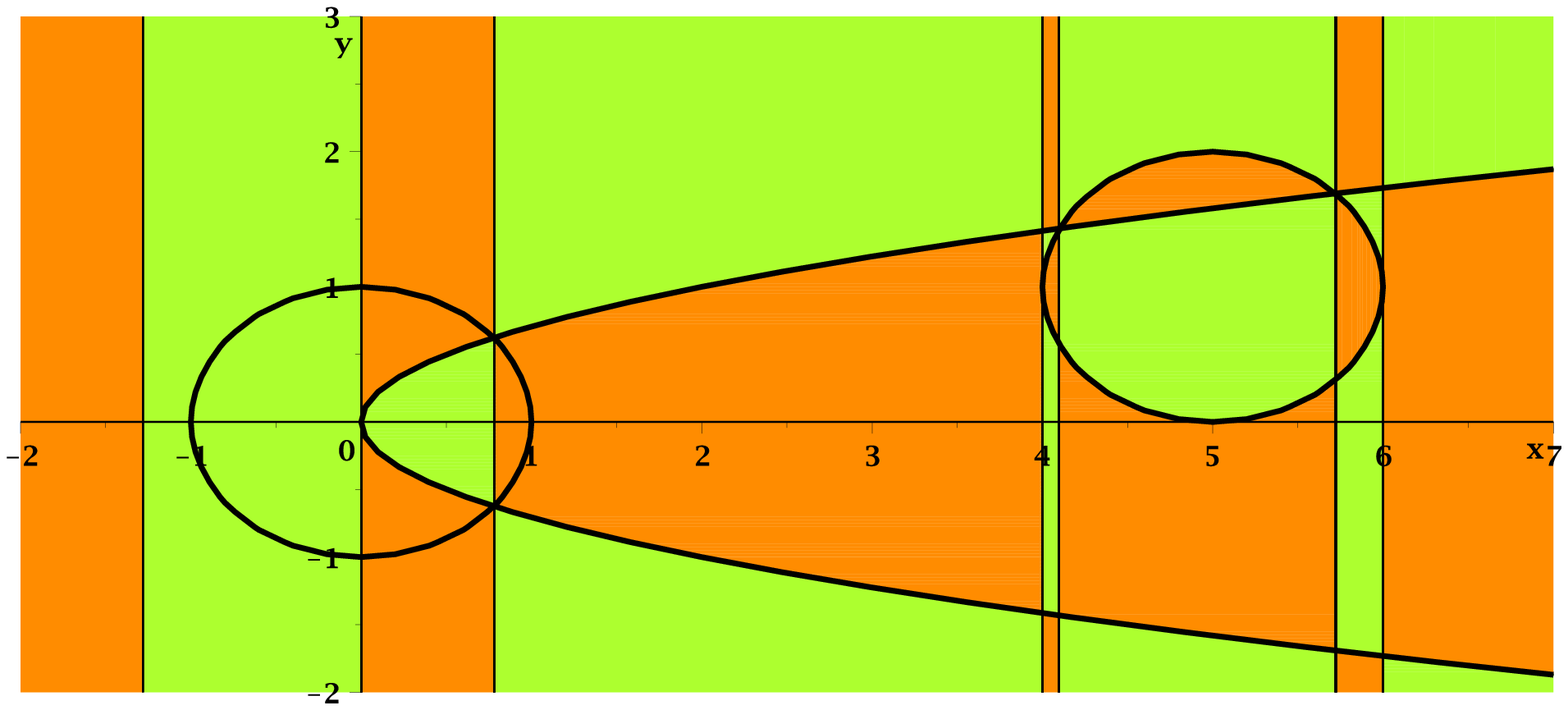}
\includegraphics[width=0.48\textwidth]{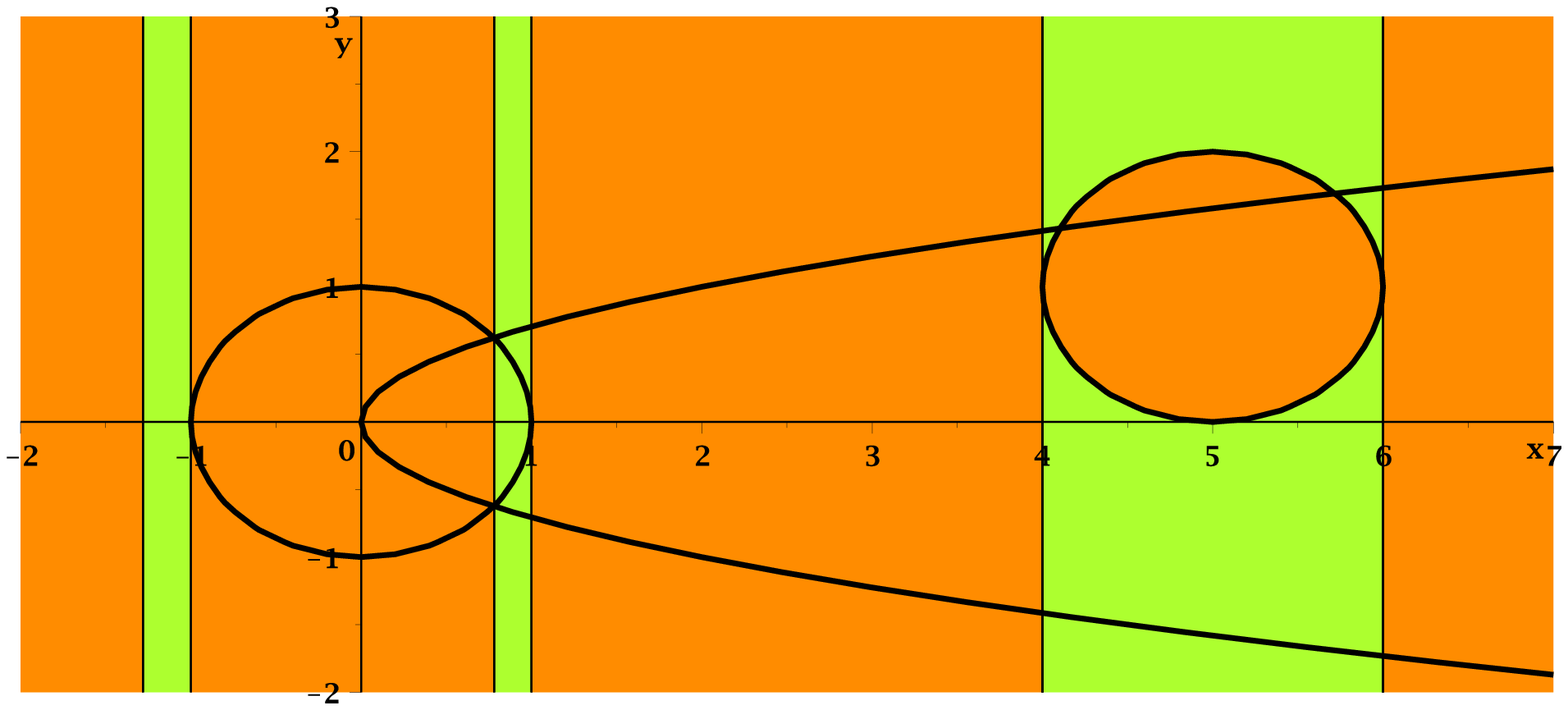}
\includegraphics[width=0.48\textwidth]{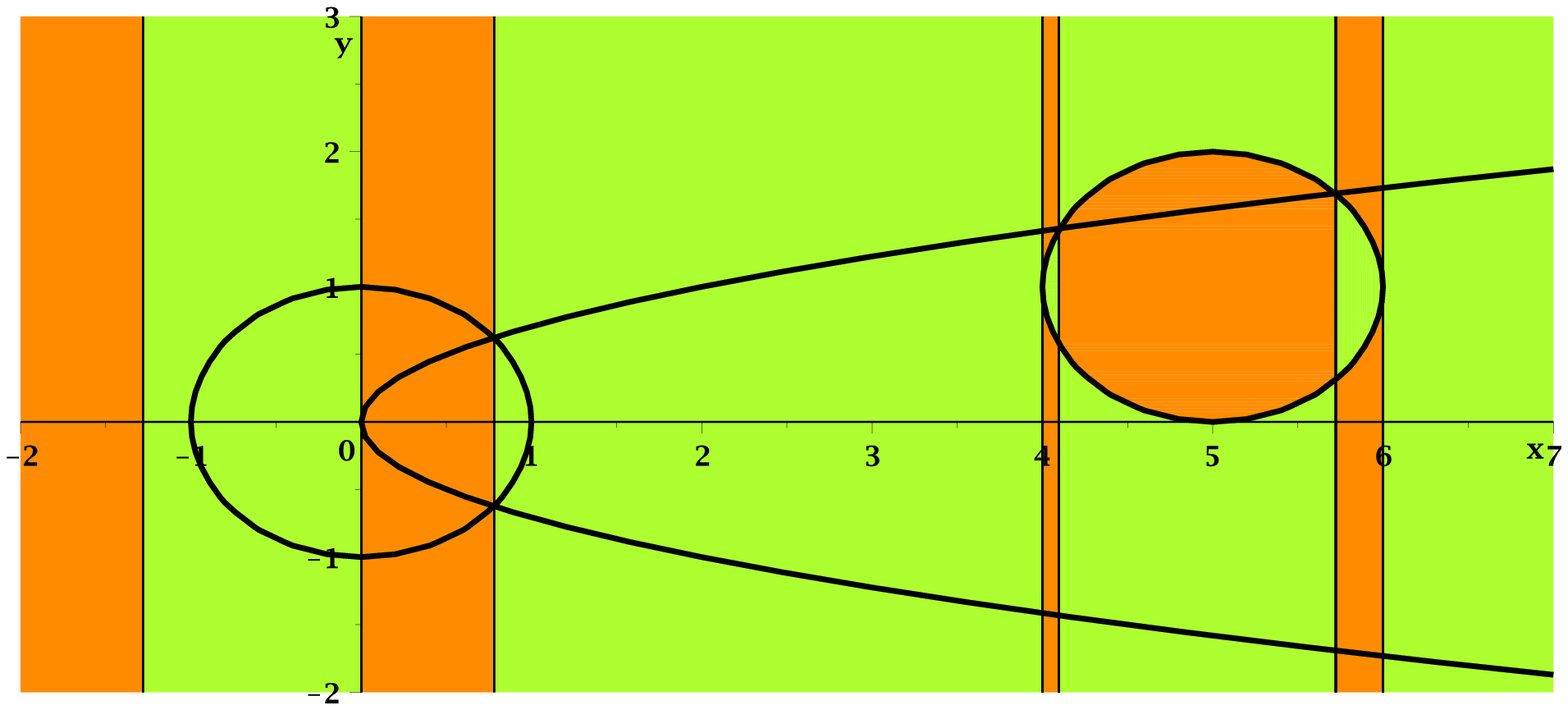}
\vspace*{-15pt}
\end{figure}

The following example illustrates why these questions matter.
\begin{example}
\label{ex:Learning}
We assume the ordering $x \prec y$ and consider
\begin{align*}
f_{1} &:= x^2+y^2-1, \qquad 
f_{2} := 2y^2-x, \qquad
f_{3} := (x-5)^2+(y-1)^2-1,  
\\
&\hspace*{0.8in} \phi_1 := f_1=0 \land f_2 = 0, \qquad
\phi_2 := f_3 = 0.
\end{align*}
The polynomials are graphed within the plots of Figure \ref{fig:Learning} (the circle on the left is $f_1$, the one on the right $f_3$ and the parabola $f_2$).  If we want to study the truth of $\phi_1$ and $\phi_2$ (or a parent formula $\phi_1 \lor \phi_2$) we need a TTICAD to take advantage of the ECs.   There are two possible answers to each of the questions above and so four possible inputs to \texttt{RC-TTICAD}.  The corresponding outputs are\footnote{All timings in this paper were obtained on a Linux desktop (3.1GHz Intel processor, 8.0Gb total memory) using \textsc{Maple 18}.}:\\
\textbf{$\bm{\phi_1 \rightarrow \phi_2}$ and $\bm{f_1 \rightarrow f_2}$:} 
37 cells in 0.095 seconds.\\
\textbf{$\bm{\phi_1 \rightarrow \phi_2}$ and $\bm{f_2 \rightarrow f_1}$:} 
81 cells in 0.118 seconds.\\
\textbf{$\bm{\phi_2 \rightarrow \phi_1}$ and $\bm{f_1 \rightarrow f_2}$:} 
25 cells in 0.087 seconds.\\
\textbf{$\bm{\phi_2 \rightarrow \phi_1}$ and $\bm{f_2 \rightarrow f_1}$:}
43 cells in 0.089 seconds.\\
The plots in Figure \ref{fig:Learning} show the two-dimensional cells in each of these TTICADs.
\end{example}

First compare the induced CADs of $\mathbb{R}^1$ (how the real line is dissected).  
Observe the following similarities in all four images:
\begin{itemize}
\item The points $\tfrac{1}{4}(-1\mp\sqrt{17})$ (approximately -1.28 and 0.78) are always identified.  
\end{itemize}
The latter is at the intersection of $f_1$ and $f_2$ and so is essential for the output to be correct as the truth of $\phi_1$ changes here.  The former is the other root of the resultant of $f_1$ and $f_2$ and so marks an intersection with complex $y$-value.  
\begin{itemize}
\item The points $4$ and $6$ are always identified.  These mark the endpoints of $f_3$, required for cylindricity and obtained as roots of the discriminant of $f_3$.  
\end{itemize}

\noindent Now we observe the differences in the induced CADs of the real line:
\begin{itemize}
\item If $f_1 \rightarrow f_2$ we identify $\pm 1$, marking the ends of the circle $f_1$.  Similarly if $f_2 \rightarrow f_1$ then we identify $0$, marking the end of the parabola $f_2$.  
\end{itemize}
These are identified as the roots of a discriminant and are included to ensure cylindricity.  If $f_1$ is processed first then $f_2=0$ is only considered when $f_1=0$.  Since their intersection is only a single value of $x$ the discriminant of $f_2$ is not required (or more accurately it is trivial).  Similarly, if we process $f_2$ first then the discriminant of $f_1$ is only calculated for two values of $x$,  where it is constant.
\begin{itemize}
\item If $f_2 \rightarrow f_1$ then we identify the two real roots of the resultant of $f_2$ and $f_3$ (approximately 4.10 and 5.72) marking the real intersection of those curves.  
\end{itemize}
If we process $f_2$ first and then $f_1$ the algorithm ensures their intersections are identified to maintain truth-table invariance.  For this example it is not necessary since when we process $f_1$ we find there are no intersections of the set where $\phi_1$ is true with $f_3$, but this was not known beforehand.  If instead $f_1 \rightarrow f_2$ since there is no intersection the extra cells are avoided.  However, note that the resultant of $f_1$ and $f_3$ is still calculated and the complex tree is split accordingly.  This may explain why the timings for the orderings with $\phi_2 \rightarrow \phi_1$ are so similar.



Finally compare the CADs of $\mathbb{R}^2$.  We see that in all four TTICADs the output is sign invariant for $f_3$, while if $\phi_1 \rightarrow \phi_2$ then the output is also sign invariant for whichever of $f_1$ and $f_2$ was processed first.
The first constraint to be processed will always be sign-invariant for the output.  The tree is initially refined into cases where its polynomial is zero or not and although these branches are split further that invariance is maintained.  Similarly, the first constraint from a formula to be processed will usually be sign-invariant in the output, but this may be avoided if a formula has more than one EC.  In this case the tree may be refined to the cases where either both are satisfied or not (as with $\phi_1$ in this example).   

\subsection{Developing a heuristic for equational constraint ordering}
\label{SUBSEC:H1}


The following propositions are illustrated by Example \ref{ex:Learning} and can be verified from the algorithm specification in \cite{BCDEMW14}.
\begin{proposition}
The output of \texttt{RC-TTICAD} is always sign-invariant with respect to the discriminant of the first EC in each formula.
\end{proposition}
Other discriminants will be calculated, but their impact is lesser.  E.g. the discriminant of the second EC in a formula will be considered modulo the first.
\begin{proposition}
The output of \texttt{RC-TTICAD} is always sign-invariant with respect to the cross-resultants of the set of first ECs in each formula.
\end{proposition}
Other resultants will be calculated.  Some of them have lesser impact, such as the resultant of the first EC in formula A with the second in formula B which is considered modulo the first EC in formula B.  Other will be considered for all constraint orderings, such as the resultant of a pair of ECs in a formula.

Considering (1) and (2) leads us to suggest minimising the following sets under some measure when making a choice about constraint ordering.
\begin{definition}
\label{def:COS}
%
For a given constraint ordering $\mathfrak{o}$ let $P$ be the set of ECs which are ordered first in each formula.  Then define the \textbf{constraint ordering set} $\mathcal{C}_{\mathfrak{o}}$ as the discriminants and cross resultants in the main variable of the ordering:
\[
\mathcal{C}_{\mathfrak{o}} := \textstyle
\Big( \bigcup_{p \in P} \big\{ \rm{disc}(p)\big\} \Big)
\cup 
\Big( \bigcup_{p,q \in P, p \neq q} \big\{ \rm{res}(p,q) \big\} \Big). 
\]
\end{definition}
For Example \ref{ex:Learning} the constraint ordering sets are
\begin{align*}
\mathcal{C}_{f_1 \rightarrow f_2} 
&= \{ {\rm disc}_{y}(f_1), {\rm disc}_{y}(f_3), {\rm res}_{y}(f_1,f_3) \} \\
&= \{ -4x^2+4, -4x^2+40x-96,  104x^2-520x+672 \}, \\
\mathcal{C}_{f_2 \rightarrow f_1} 
&= \{ {\rm disc}_{y}(f_2), {\rm disc}_{y}(f_3), {\rm res}_{y}(h,f_2) \} \\
&= \big\{ 8x, -4x^2+40x-96, 4x^4-76x^3+561x^2-1908x+2500 \big\}.
\end{align*}
A natural way to measure these would be to compare \texttt{sotd}s, the \emph{sum of total degrees of each monomial in each polynomial}, since this was shown to help with other CAD choices \cite{DSS04, BDEW13}.  For Example \ref{ex:Learning} the sets above have sotd $8$ and $14$ respectively and thus the ordering $f_1 \rightarrow f_2$ is suggested.  Regardless of which formula is processed first, this is the better choice.  However, the following example demonstrates that sotd may not be so suitable in general.

\begin{example}
\label{ex:2014Ellipse}
\textbf{[2014 $\bm{x}$-axis ellipse problem]}
A well studied test problem for CAD is the \emph{$x$-axis ellipse problem} defined in \cite{AM88} and specialising a problem in \cite{Kahan75}.  They concern an ellipse and seek to determine for what values of its parameters (the principal semi-axes and centre) it lies within the unit circle. 

We propose a new problem, inspired by the original but requiring multiple formulae and hence a TTICAD.  Suppose the ellipse is centred at $(c,0)$ with principal semi-axes $a \in (0,2)$ and $b=1$.  The problem is to determine for what values of $(c,a)$ the ellipse intersects either of a pair of unit circles, centred on the $x$-axis at $\pm 2$.  Define the polynomials
\begin{align*}
f_1 := (x-2)^2 + y^2 - 1, \quad 
f_2 := (x+2)^2 + y^2 - 1, \quad
h := (x-c)^2 + a^2y^2 - a^2.
\end{align*}
Then we seek to eliminate the quantifiers from $\Phi := (\exists y) (\exists x) \phi_1 \lor \phi_2$ where
\begin{align*}
\phi_1 &:= (f_1=0 \land h=0 \land a>0 \land a<2), \, 
\phi_2 := (f_2=0 \land h=0 \land a>0 \land a<2).
\end{align*}
The problem can be solved using a TTICAD for $\phi_1$ and $\phi_2$.  We assume variable ordering $y \succ x \succ a \succ c$.  There are eight possible constraint orderings for \texttt{RC-TTICAD} as listed in Table \ref{tab:ellipse}.  The best choice is to process $h$ first in each formula (then the formula ordering makes no difference) which is  logical since $h$ has no intersections with itself to identify. 
However, using sotd as a measure on the constraint ordering set will lead us to select the very worst ordering.  Consider the constraint ordering sets for these two cases:
\begin{align*}
\mathcal{C}_{f_1 \rightarrow h, f_2 \rightarrow h} 
&= \{ {\rm disc}_{y}(f_1), {\rm disc}_{y}(f_2), {\rm res}_{y}(f_1,f_2) \} \\
&= \{ -4x^2+16x-12, -4x^2-16x-12, 64x^2 \}, \\
\mathcal{C}_{h \rightarrow f_1, h \rightarrow f_2} 
&= \{ {\rm disc}_{y}(h), {\rm res}_{y}(h,h) \} 
= \big\{ 4a^2(a^2-c^2+2cx-x^2), 0 \big\}.
\end{align*}
Although the first has three polynomials and the second only one, this one has a higher sotd than the first three summed.  This is because only $h$ contained the parameters $(a,c)$ while $f_1$ and $f_2$ did not, but their presence was not as significant as the complexity in $x$ alone.  A more suitable measure would be the sum of degrees in $x$ alone (shown in the final column in Table \ref{tab:ellipse}) in which the first has 6 and the second only 2.
\end{example}  

\begin{table}[t]
\vspace*{-15pt}
\caption{Details on the TTICADs that can be built using \texttt{RC-TTICAD} for Example \ref{ex:2014Ellipse}.
}
\label{tab:ellipse}
\centering
\begin{tabular}{|ccc|cc|cc|}
\hline
\multicolumn{3}{|c|}{\textbf{Constraint Ordering} $\bm{\mathfrak{o}}$} & \multicolumn{2}{c|}{\textbf{TTICAD}} 
& \multicolumn{2}{c|}{ $\bm{\mathcal{C}_{\mathfrak{o}}}$ } \\
\, Formula order \,&\, $\phi_1$ order \,&\, $\phi_2$ order    \,&\, Cells \,&\, Time (sec)                     \,&\, sotd \,&\, deg \\
\hline 
$\phi_1 \rightarrow \phi_2$ & $h \rightarrow f_1$ & $h \rightarrow f_2$ & 24545  &   86.082 & 16   & 2 \\
$\phi_1 \rightarrow \phi_2$ & $h \rightarrow f_1$ & $f_2 \rightarrow h$ & 73849  &  499.595 & 114  & 8 \\
$\phi_1 \rightarrow \phi_2$ & $f_1 \rightarrow h$ & $h \rightarrow f_2$ & 67365  &  414.314 & 114  & 8 \\
$\phi_1 \rightarrow \phi_2$ & $f_1 \rightarrow h$ & $f_2 \rightarrow h$ & 105045 & 1091.918 & 8    & 6 \\
$\phi_2 \rightarrow \phi_1$ & $h \rightarrow f_1$ & $h \rightarrow f_2$ & 24545  &   87.378 & 16   & 2 \\
$\phi_2 \rightarrow \phi_1$ & $h \rightarrow f_1$ & $f_2 \rightarrow h$ & 67365  &  401.598 & 114  & 8 \\
$\phi_2 \rightarrow \phi_1$ & $f_1 \rightarrow h$ & $h \rightarrow f_2$ & 73849  &  494.888 & 114  & 8 \\
$\phi_2 \rightarrow \phi_1$ & $f_1 \rightarrow h$ & $f_2 \rightarrow h$ & 105045 & 1075.568 & 8    & 6 \\
\hline
\end{tabular}
\vspace*{-15pt}
\end{table}

\begin{remark}
It is not actually surprising that sotd is inappropriate here while working well in \cite{DSS04, BDEW13}.  In those studies sotd was measuring projection sets (either the whole set or at one stage in the projection) while here we are measuring only the subset which changes the most with the ordering.  Sotd is principally a measure of sparseness.  Sparseness of the entire projection set indicates less complexity while sparseness of one level is likely to lead to sparseness at the next.  However, the constraint ordering set being sparse does not indicate that the other polynomials involved at that stage or subsequent ones will be.
\end{remark}

\begin{heuristic}
\label{def:H1}
Define the \textbf{EC ordering heuristic} as selecting the first EC to be processed in each formula such that the corresponding constraint ordering set has lowest sum of degrees of the polynomials within (all taken in the second variable of the ordering).
\end{heuristic}
Heuristic \ref{def:H1} follows from the analysis above and we evaluate it in Section \ref{SEC:Results}.  We can already see three apparent shortcomings:
\begin{enumerate}[(i)]
\item How to break ties if the sum of degrees are the same?
\item What to do if the complex geometry is different to the real geometry?
\item How to order remaining equational constraints?
\end{enumerate}
One answer to (i) is to break ties with sotd.  A tie with Heuristic \ref{def:H1} is a good indication that the complex geometry in the highest dimension is of the same complexity and so further discrimination will require lower dimensional components.  In fact, these are also needed to address (iii).  Suppose a formula contained three ECs and we had determined which to process first.  Then the choice of which is second means comparing the resultant of the first with each of the others modulo the first.  In our experience such formulae tend to give similar output for the different orderings due to the simplifications in the tree so many ECs offer. 

\begin{figure}[t]
\vspace*{-15pt}
\caption{Visualisations of two TTICADs built using \texttt{RC-TTICAD} for Example \ref{ex:LearningShifted}.  They both have $\phi_2 \rightarrow \phi_1$, with the first having $f_1 \rightarrow f_2$ and the second $f_2 \rightarrow f_1$.
}
\label{fig:LearningShifted}
\centering
\includegraphics[width=0.48\textwidth]{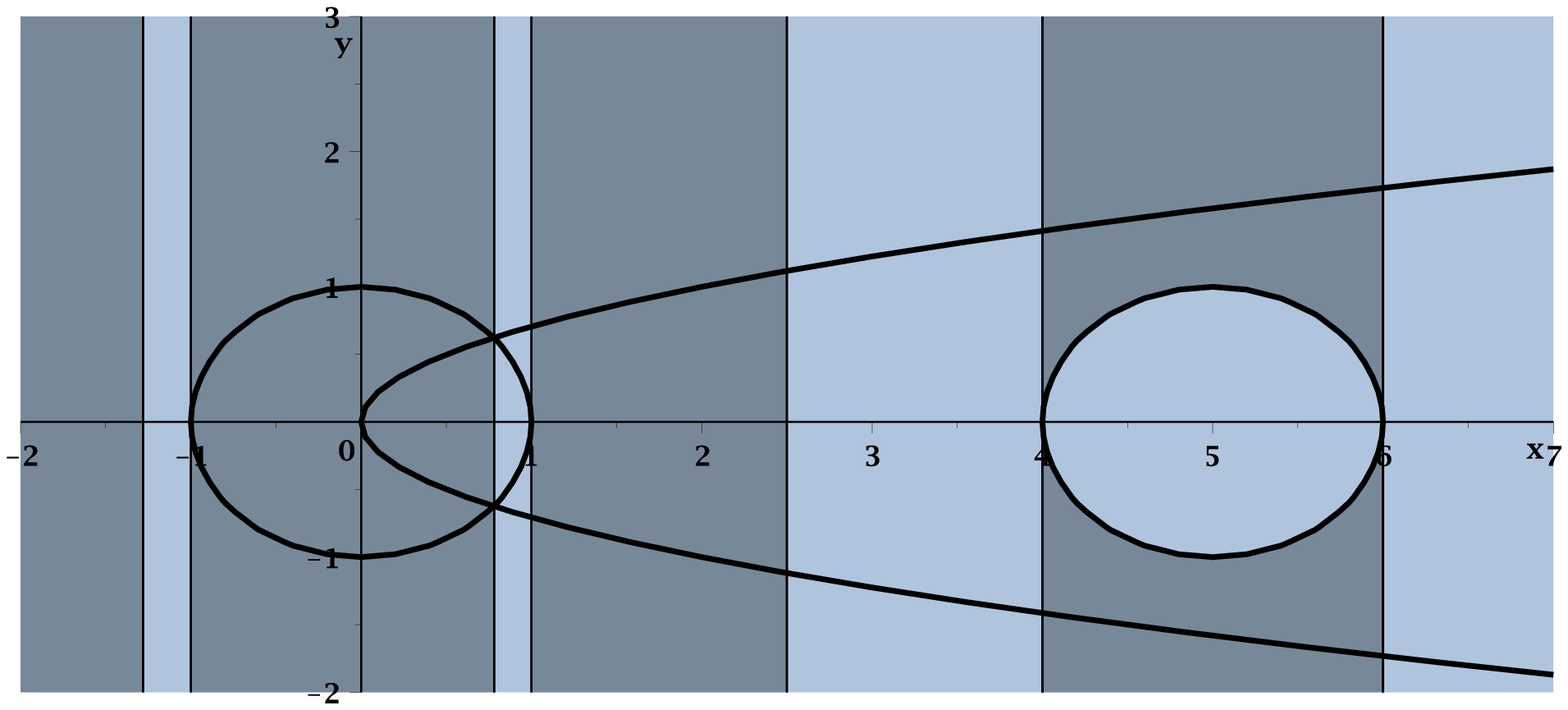}
\includegraphics[width=0.48\textwidth]{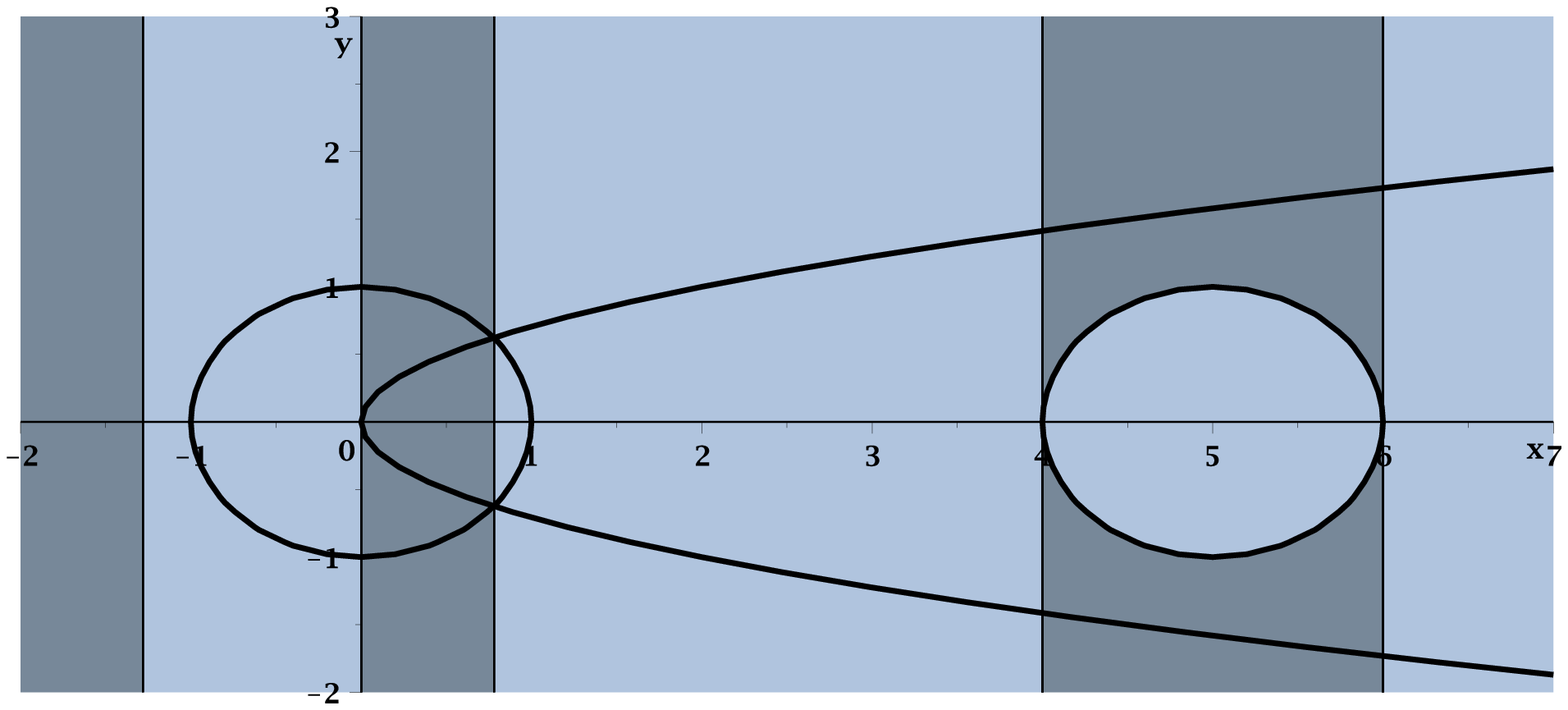}
\vspace*{-15pt}
\end{figure}

Heuristic \ref{def:H1} can be misled as suggested by (ii) and demonstrated next.
\begin{example}
\label{ex:LearningShifted}
Consider the polynomials and formulae from Example \ref{ex:Learning} but with $f_2$ and $g_2$ shifted under $y \mapsto y+1$.  The possible outputs from \texttt{RC-TTICAD} are:
\\
\textbf{ $\bm{\phi_1 \rightarrow \phi_2}$ and $\bm{f_1 \rightarrow f_2}$:} 
39 cells in 0.094 seconds. \\
\textbf{ $\bm{\phi_1 \rightarrow \phi_2}$ and $\bm{f_2 \rightarrow f_1}$:} 
49 cells in 0.081 seconds. \\
\textbf{ $\bm{\phi_2 \rightarrow \phi_1}$ and $\bm{f_1 \rightarrow f_2}$:} 
27 cells in 0.077 seconds. \\
\textbf{ $\bm{\phi_2 \rightarrow \phi_1}$ and $\bm{f_2 \rightarrow f_1}$:} 
23 cells in 0.073 seconds. \\
Since $f_2$ no longer intersects $h$ the best choice is the fourth instead of the third.
Figure \ref{fig:LearningShifted} compares these two TTICADs.  The only difference now is whether the endpoints of the left circle or the parabola are identified.  Since the parabola has only one endpoint it becomes the better choice.  However, the constraint ordering set has the same degree in $x$ or sotd and so still suggests $f_1 \rightarrow f_2$.  
\end{example}

Heuristic \ref{def:H1} is misled here because the degree is a measure only of the behaviour in complex space, which did not change significantly between the examples.
In \cite{BDEW13} we demonstrated similar issues for CAD (and TTICAD) by projection and lifting.  There we devised an alternative heuristic: the \emph{number of distinct real roots of the univariate polynomials} (ndrr) which meant essentially comparing the induced CADs of the real line.  However, \texttt{RC-TTICAD} does not start by directly computing all polynomials involved in the computation (the projection phase).  Example \ref{ex:LearningShifted} is in only two dimensions and so the ndrr could easily be applied to the univariate constraint ordering sets to suggest the best ordering.  But for higher dimensional examples it is not so clear what or how to measure.  Further, the complex geometry does have a direct effect on \texttt{RC-TTICAD} not present in the projection and lifting algorithms since we first build a CCD.

\subsection{Developing a heuristic for formulae ordering}
\label{SUBSEC:H2}

Heuristic \ref{def:H1} helps with ordering ECs in formulae but not how to order the formulae themselves.  In Example \ref{ex:Learning} the main difference between formulae orderings was which polynomial is ensured sign-invariant in the output. In Example \ref{ex:Learning} there was a clear choice to process $\phi_2$ first since its sole EC would be sign-invariant regardless.  
In general we advise \textbf{placing a formula with only one EC first}.  

\begin{remark}
In fact, the analysis so far suggests that the best choice would be to process a non-EC from a formula with no ECs first.  This is because all the non-ECs in such a formula will always be sign-invariant in the output and so dealing with them first would occur no cost but possibly allow savings from another formulae with multiple ECs.  The algorithm as specified in \cite{BCDEMW14} does not allow this but we intend to investigate this possibility in future work. 
\end{remark}

We now seek a heuristic to help with formulae ordering when no obvious choice is available.  Ideally, we require an (efficient) measure of how large the (real) projection is of a polynomial out of its main variable, but such a measure is not clear to us.  Instead we explore an alternative approach.  As discussed, the CAD algorithms based on regular chains technology first build a CCD before refining to a CAD.  It has been observed that the refinement to real space usually takes the most time (involving real root isolation), but that the timings of the two stages are correlated.   Hence, we consider building the CCD first for multiple orderings and then choosing the smallest one.
\begin{heuristic}
\label{def:H2}
Define the \textbf{CCD size heuristic} as selecting a constraint ordering by constructing the CCD for each, extracting the set of polynomials used in each tree, and choosing the one to refine to a CAD whose set has the lowest sum of degree of the polynomials within (each taken in the main variable of that polynomial).
\end{heuristic}
We evaluate this heuristic in the next section. It clearly requires far more computation than Heuristic \ref{def:H1} and so the relative costs will have to be considered.  This leads us to suggest a third heuristic combining the approaches.

\begin{heuristic}
\label{def:H3}
Define the \textbf{constraint ordering heuristic} as using Heuristic \ref{def:H1} to suggest the best subset of constraint orderings and then having Heuristic \ref{def:H2} pick from these, splitting any further ties by picking lexicographically.  
\end{heuristic}

\section{Evaluating the heuristics}
\label{SEC:Results}

\subsection{Experiments and data}
\label{SUBSEC:Data}

We tested the effectiveness of the heuristic using 100 random systems of the form
\[
\phi_1 := (f_1=0 \land f_2=0 \land g_1>0),  \qquad
\phi_2 := (f_3=0 \land f_4=0 \land g_2>0).
\]
The polynomials were randomly generated using \textsc{Maple}'s \texttt{randpoly} command as sparse polynomials in the ordered variables $x \prec y \prec z$ with maximum degree 3 and integer coefficients .  Each problem has three questions of constraint ordering: 

\vspace*{0.03cm}
\noindent \, - Process $\phi_1$ first or $\phi_2$?  \quad
- Process $f_1$ first or $f_2$? \quad
- Process $f_3$ first or $f_4$?

Hence each problem has eight possible orderings.  We build TTICADs using \texttt{RC-TTICAD} for each ordering and compare the number of cells and computation time.  We set a time limit of 40 minutes per problem (so an average of 5 minutes per CAD) in which 92 of the problems could be studied.  The average CAD had 2018.3 cells and was computed in 36.5 seconds, but there were several outliers bringing the average up.  The median values were 1554.1 cells and 6.1 seconds.

For each problem we considered how Heuristics \ref{def:H1}, \ref{def:H2} and \ref{def:H3} performed.  We start by comparing cell counts in Table \ref{tab:ResultsC}.  For each problem we calculated:
\begin{enumerate}[(a)]
\item The average cell count of the 8 TTICADs computed.
\item The average cell count of the TTICADs selected by each heuristic (but note that Heuristic \ref{def:H3} always selects only one).  
\item The average saving from using each heuristic, computed as (a)$-$(b).
\item The average percentage saving to the cell count, calculated as 100(c)/(a).
\end{enumerate}
The figures in Table \ref{tab:ResultsC} show the values of (b)$-$(d) for each heuristic, averaged over the 92 problems.  To compare timings we calculated (a$'$)$-$(d$'$) as the equivalent of (a)$-$(d) for timings.  Then for each problem we also calculated:
\begin{enumerate}[(a$'$)] 
\setcounter{enumi}{4}
\item The time taken to run each heuristic.
\item The net saving calculated as (a$'$)$-$(b$'$)$-$(e$'$).
\item The net percentage saving calculated as 100(f$'$)/(a$'$).
\end{enumerate}
Table \ref{tab:ResultsT} shows the values of (b$'$)$-$(d$'$),(f$'$),(g$'$) averaged over the 92 problems.

\begin{table}[t]
\vspace*{-15pt}
\caption{How the CADs selected by the heuristics compare on cell count.}
\label{tab:ResultsC}
\centering
\begin{tabular}{|l||c|c|c|}
\hline
\, \textbf{Heuristic} \,&\, Cell Count \,&\, Saving \,&\, \% Saving \, \\  
\hline
\, Heuristic \ref{def:H1} & 1589.67                & 428.61             & 26.73               \\
\, Heuristic \ref{def:H2} & 1209.10                & 809.18             & 47.70               \\
\, Heuristic \ref{def:H3} & 1307.63                & 710.65             & 40.97               \\
\hline
\end{tabular}
\end{table}

\begin{table}[t]
\caption{How the CADs selected by the heuristics compare on timings (in seconds).}
\label{tab:ResultsT}
\centering
\begin{tabular}{|l||c|c|c|c|c|}
\hline
\,\textbf{Heuristic}\,&\,Timing\,&\,Saving\,&\,\% Saving\,&\,Net Saving\,&\,\% Net Saving \, \\  
\hline
\, Heuristic \ref{def:H1}\, & 14.48  & 22.02    & 37.17       & 22.01        & 37.12          \\
\, Heuristic \ref{def:H2}\, & 9.02   & 27.47    & 49.45       & -150.59      & -215.31        \\
\, Heuristic \ref{def:H3}\, & 9.42   & 27.08    & 43.84       & -20.02       & 0.77           \\
\hline
\end{tabular}
\vspace*{-15pt}
\end{table}

Tables \ref{tab:ResultsSpreadC} and \ref{tab:ResultsSpreadT} shows where the selections made by each heuristic lie on the spread of possible outputs (where 1 is the CAD with the smallest and 8 the one with the biggest).  In the event of two CADs having the same value a selection is recorded with the higher ranking.  
Since Heuristics \ref{def:H1} and \ref{def:H2} can pick more than one ordering we also display the figures as percentages.  So for example, a selection by the first heuristic was the very best ordering 24\% of the time.

\begin{table}[t]
\vspace*{-15pt}
\caption{How the heuristics selections rank out of the possible CADs for cell counts.}
\label{tab:ResultsSpreadC}
\centering
\begin{tabular}{|ll||c|c|c|c|c|c|c|c||c|}
\hline 
\, \textbf{Heuristic}    &  & 1  & 2  & 3  & 4  & 5  & 6  & 7  & 8 & Total \\
\hline
\hline
\multirow{2}{*}{\, \textbf{Heuristic \ref{def:H1}} \,}  
& \# & 60    & 46    & 44    & 26    & 21   & 17   & 17   & 15   & 246    \\
& \% & 24.39 & 18.70 & 17.89 & 10.57 & 8.54 & 6.91 & 6.91 & 6.10 & 100.01 \\
\hline
\multirow{2}{*}{\, \textbf{Heuristic \ref{def:H2}} \,}  
& \# & 55    & 19    & 12    & 5    & 5    & 2    & 0 & 0 & 98    \\
& \% & 56.12 & 19.39 & 12.24 & 5.10 & 5.10 & 2.04 & 0 & 0 & 99.99 \\
\hline
\multirow{2}{*}{\, \textbf{Heuristic \ref{def:H3}} \,} 
& \# & 44    & 22    & 7    & 6    & 4    & 4    & 3    & 2    & 92     \\
& \% & 47.83 & 23.91 & 7.61 & 6.52 & 4.35 & 4.35 & 3.26 & 2.17 & 100.00 \\
\hline
\end{tabular}
\end{table}

\begin{table}[t]
\caption{How the heuristics selections rank out of the possible CADs for timings.}
\label{tab:ResultsSpreadT}
\centering
\begin{tabular}{|ll||c|c|c|c|c|c|c|c||c|}
\hline 
\, \textbf{Heuristic}    &  & 1  & 2  & 3  & 4  & 5  & 6  & 7  & 8 & Total \\
\hline
\hline
\multirow{2}{*}{\, \textbf{Heuristic \ref{def:H1}} \,}  
& \# & 64    & 51    & 33    & 24   & 23   & 25    & 13   & 13   & 246    \\
& \% & 26.02 & 20.73 & 13.41 & 9.76 & 9.35 & 10.16 & 5.29 & 5.29 & 100.01 \\
\hline
\multirow{2}{*}{\, \textbf{Heuristic \ref{def:H2}} \,}  
& \# & 44    & 29     & 12    & 4    & 1    & 4    & 1    & 3    & 98    \\
& \% & 44.90 &  29.59 & 12.24 & 4.08 & 1.02 & 4.08 & 1.02 & 3.06 & 99.99 \\
\hline
\multirow{2}{*}{\, \textbf{Heuristic \ref{def:H3}} \,} 
& \# & 37    & 26    & 9    & 4    & 1    & 6    & 7    & 2    & 92     \\
& \% & 40.22 & 28.26 & 9.78 & 4.35 & 1.09 & 6.52 & 7.61 & 2.17 & 100.00 \\
\hline
\end{tabular}
\vspace*{-15pt}
\end{table}

\subsection{Interpreting the results}

First we observe that all three heuristics will on average make selections on constraint ordering with substantially lower cell counts and timings than the problem average.  As expected, the selections by Heuristic \ref{def:H2} are on average better than those by Heuristic \ref{def:H1}.  In fact, the measure used by Heuristic \ref{def:H2} seems to be correlated to both the cell counts and timings in the final TTICAD.  

\begin{figure}[t]
\vspace*{-15pt}
\caption{These plots compare the measures used by the heuristics with the CADs computed in Section \ref{SEC:Results}.  The plots on the left have cell count on the vertical axis, and those on the right timings.  
The horizontal axes have the sum of degrees of polynomials in a set.  On the top this is the constraint ordering set and on the bottom the polynomials in the CCD.
All values are scaled to the problem they originate from.
}
\label{fig:Correlation}
\centering
\includegraphics[width=0.45\textwidth]{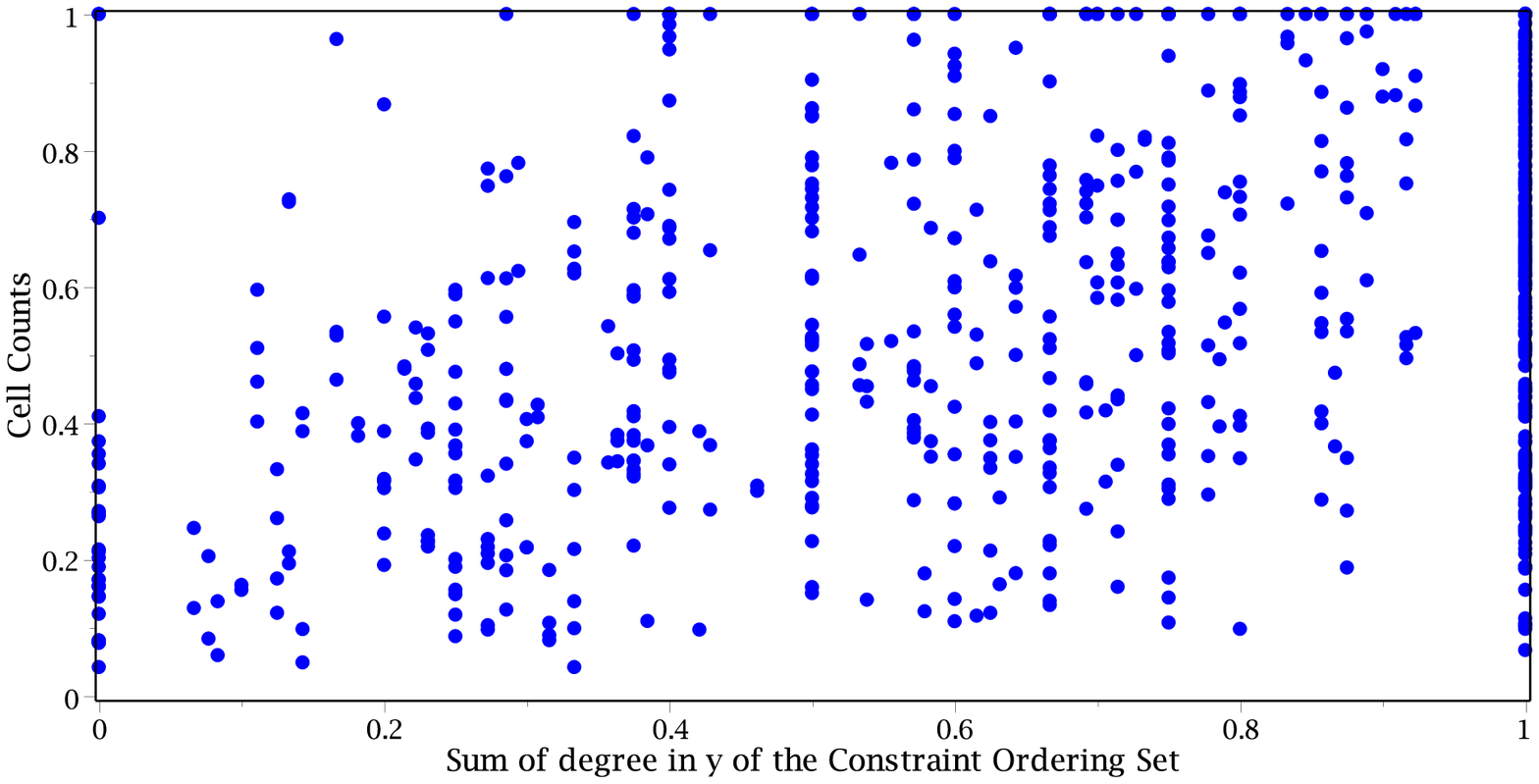}
\includegraphics[width=0.45\textwidth]{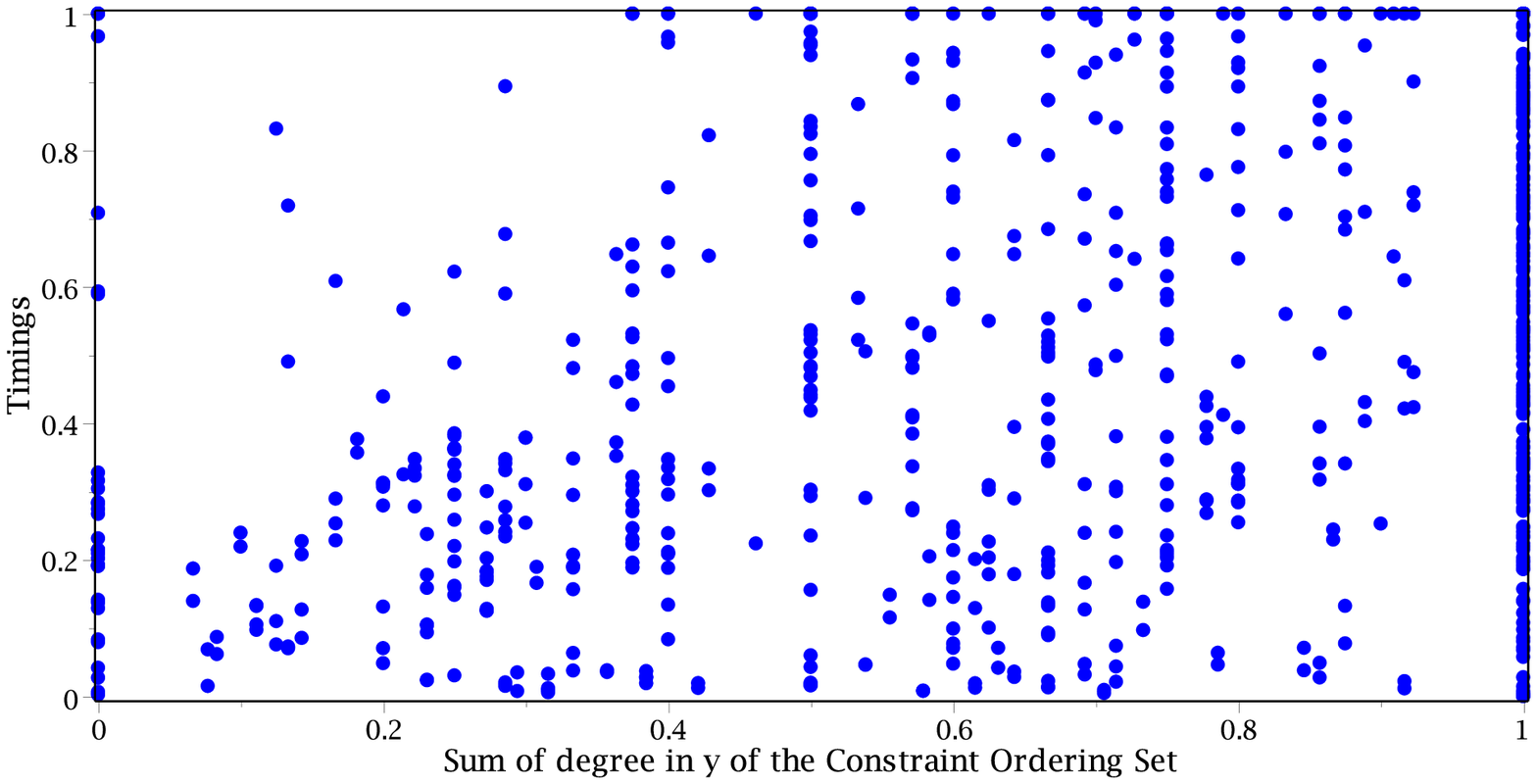}
\includegraphics[width=0.45\textwidth]{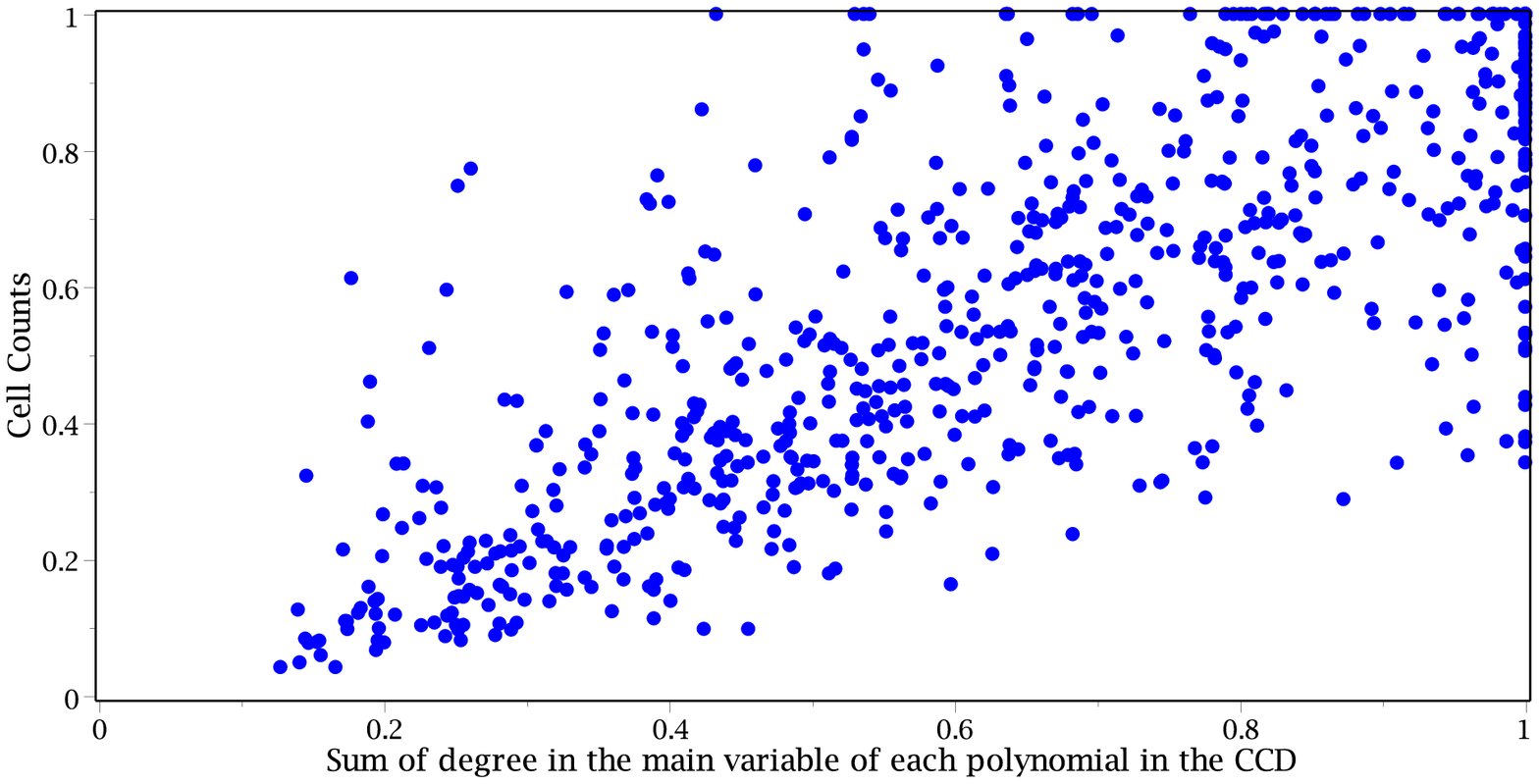}
\includegraphics[width=0.45\textwidth]{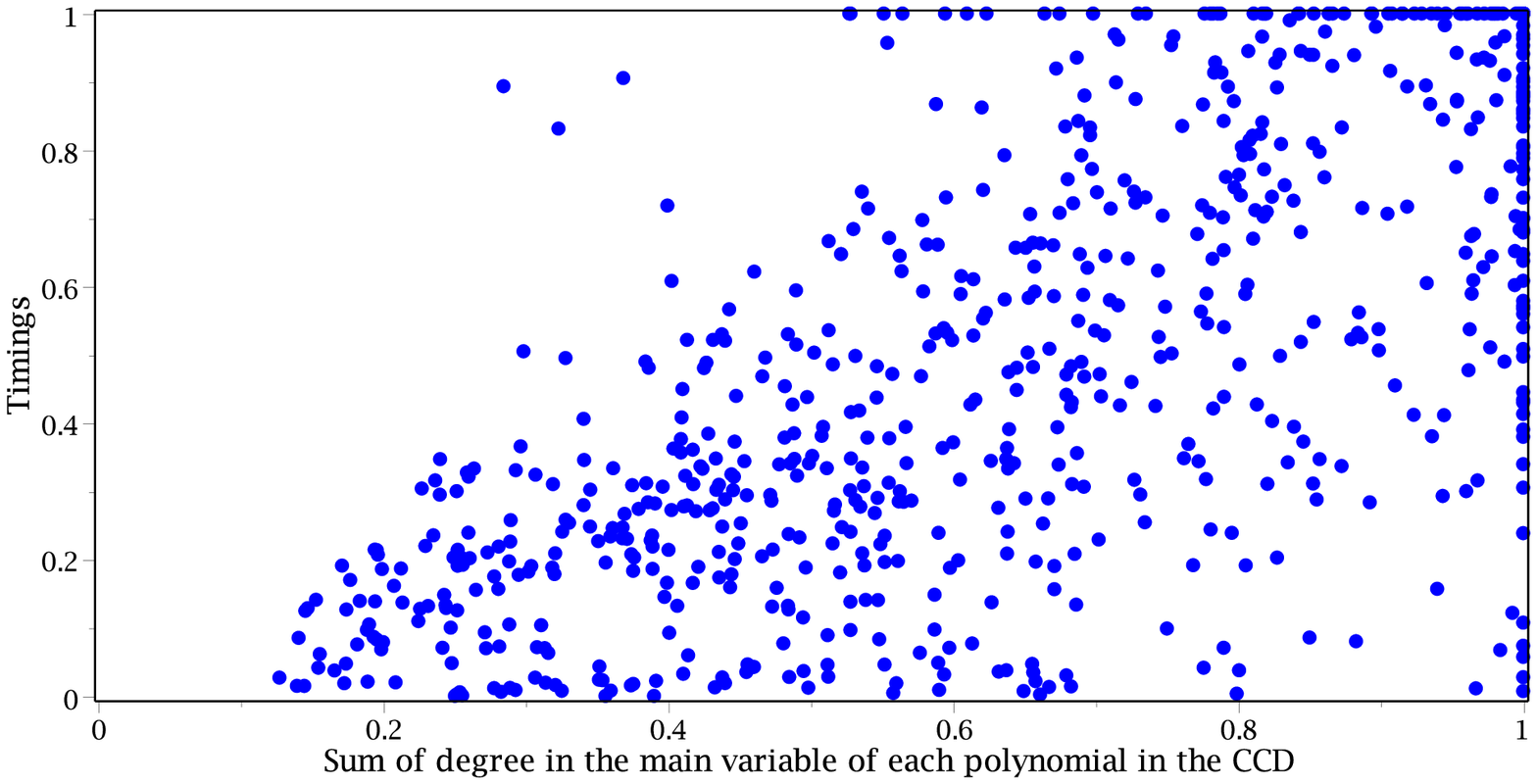}
\vspace*{-10pt}
\end{figure}

To consider the correlation we recorded the value of the measures used by Heuristics \ref{def:H1} and \ref{def:H2} and paired these with the corresponding cell counts and timings.  This was done for each CAD computed (not just those the heuristics selected).
The values were scaled by the maximum for each problem.  (Note that Heuristic 3 did not have its own measure, it was a combination of the two.) 
Figure \ref{fig:Correlation} shows the plots of these data.  The correlation coefficients for the first measure were 0.43 with cell count and 0.40 with timing, while for the second measure 0.78 and 0.68.
Since the second measure essentially completes the first part of the algorithm the correlation may not seem surprising.  However, it suggests that on average the geometry of the real and complex decomposition are more closely linked than previously thought.  This will be investigated in future work.

Although Heuristic \ref{def:H2} makes good selections, its cost is usually larger than any potential time savings (roughly 6 times larger on average).  Further, this cost will rise with the number of orderings far quicker than the cost of the others.  We note that the magnitude of this cost is inflated by the outliers, the average cost being 178.06 seconds while the median only 13.43.  Heuristic \ref{def:H1} is far cheaper, essentially zero. Although the savings were not as high they were still significant, with most selections being among the best.  We recommend Heuristic \ref{def:H1} as a cheap test to use before running the algorithm and it will likely become part of the default implementation.

The results for Heuristic \ref{def:H3} which used a mixture of the approaches are particularly interesting.  It offers substantially more savings than Heuristic \ref{def:H1}, almost achieving those those Heuristic \ref{def:H2}  but its cost is on average 47.10 seconds (with a median value of 7.55), far less than those of Heuristic \ref{def:H2}.  On average Heuristic \ref{def:H3} took more time in total to compute than its time savings, but when we consider the percentage saving the average is (just) positive.  This is not a mistake: the results are as stated because a number of outliers had a very high cost while for most examples the cost was significantly less than the savings.  

We can see situations where all three heuristics could be of use:
\\
\textbf{Use Heuristic \ref{def:H1}} if lowest computation time is prioritised, for example if many CADs must be computed or this is just a small step in a larger calculation. \\
\textbf{Use Heuristic \ref{def:H2}} if lowest cell count is prioritised, for example if only one CAD must be computed but then much work will be performed with its cells. \\
\textbf{Use Heuristic \ref{def:H3}} for a mixed approach, for example if a low cell count is required but the problem size makes Heuristic \ref{def:H2} infeasible.

\section{Other issues of problem formulation}
\label{SEC:Other}

For the original TTICAD algorithm (by projection and lifting) \cite{BDEMW13} the ordering of the constraints is not important, but other issues are, as investigated in \cite{BDEW13}.  We revisit two of those issues to see if further consideration is needed for \texttt{RC-TTICAD}.

\subsection{Equational constraint designation}  
\label{SUBSEC:Designation}

The TTICAD algorithm by projection and lifting \cite{BDEMW13} made use of a single \emph{designated} EC per formula (any others were treated the same as non-ECs).  Indeed, this projection operator generalised the one in \cite{McCallum1999} for a formula with one EC and in either case the user needs to make this designation before running the algorithm.  
\texttt{RC-TTICAD} \cite{BCDEMW14} (and the algorithm in \cite{CM12b}) can take advantage of more than one EC per formula and so the user only needs to choose the order they are used in.  We observe that the choice of which EC to process first is analogous to choosing which to designate.  For example, consider two formulae of the form
\[
\phi_i := f_1 = 0 \land f_2 = 0 \land g_1 < 0, \quad \phi_2 := f_3 = 0 \land g_2 = 0.
\]
Then the resultants and discriminants that must be calculated for the first projection phase using the operator in \cite{McCallum1999} are
\[
\{ \mbox{res}(f_i,f_j), \mbox{res}(f_i,g_1), \mbox{res}(f_i, f_3), \mbox{disc}(f_i), \mbox{disc}(f_3) \}
\] 
if $f_i$ is designated and $f_j$ not.
All polynomials from the constraint ordering set are contained here, as can be shown for the general case.  A good choice of designation for the projection and lifting algorithm is hence likely to correspond to a good choice of which EC from a formula to process first in the regular chains algorithm.  We hope to investigate this further in the future.

\subsection{Composing sub-formulae}
\label{SUBSEC:SubForm}

Consider 
$
\Phi := (f_1=0 \land \psi_1) \lor (f_2=0 \land \psi_2).
$
where $\psi_1, \psi_2$ are conjunctions.
We seek a truth-invariant CAD for $\Phi$ but neither of the equations are ECs (at least not explicitly without knowledge of $\psi_1$ and $\psi_2$).  One option would be to use $f_1f_2 = 0$ as an EC (this is logically implied by $\Phi$).  Another option is to define
\[
\phi_1 := f_1=0 \land \psi_1, \qquad \phi_2 := f_2=0 \land \psi_2
\]
and construct a TTICAD for them (any TTICAD for $\phi_1,\phi_2$ is truth-invariant for $\Phi$).  For the projection and lifting algorithms the second approach is preferable as the projection set for the latter is contained in the former.  \texttt{RC-TTICAD} requires as input semi-algebraic systems each representing a single conjunctive formula.  Hence here there is not even an analogue of the former approach.  

However, there was a similar question posed in \cite[Section 4]{BDEW13} which we now investigate in reference to \texttt{RC-TTICAD}.  Consider the single conjunctive formulae,
$
\hat{\Phi} := f_1=0 \land \psi_1 \land f_2=0 \land \psi_2,
$
where $\psi_1, \psi_2$ are again conjunctions.  We could build a CAD for $\hat{\Phi}$ or a TTICAD for $\phi_1,\phi_2$ as above.  While the projection set for the latter is in general smaller, the following example gives an exception.  

\begin{figure}[t]
\vspace*{-15pt}
\caption{Visualisations of CADs that can be built for Example \ref{ex:Form}.
}
\label{fig:Form}
\centering
\includegraphics[width=0.243\textwidth]{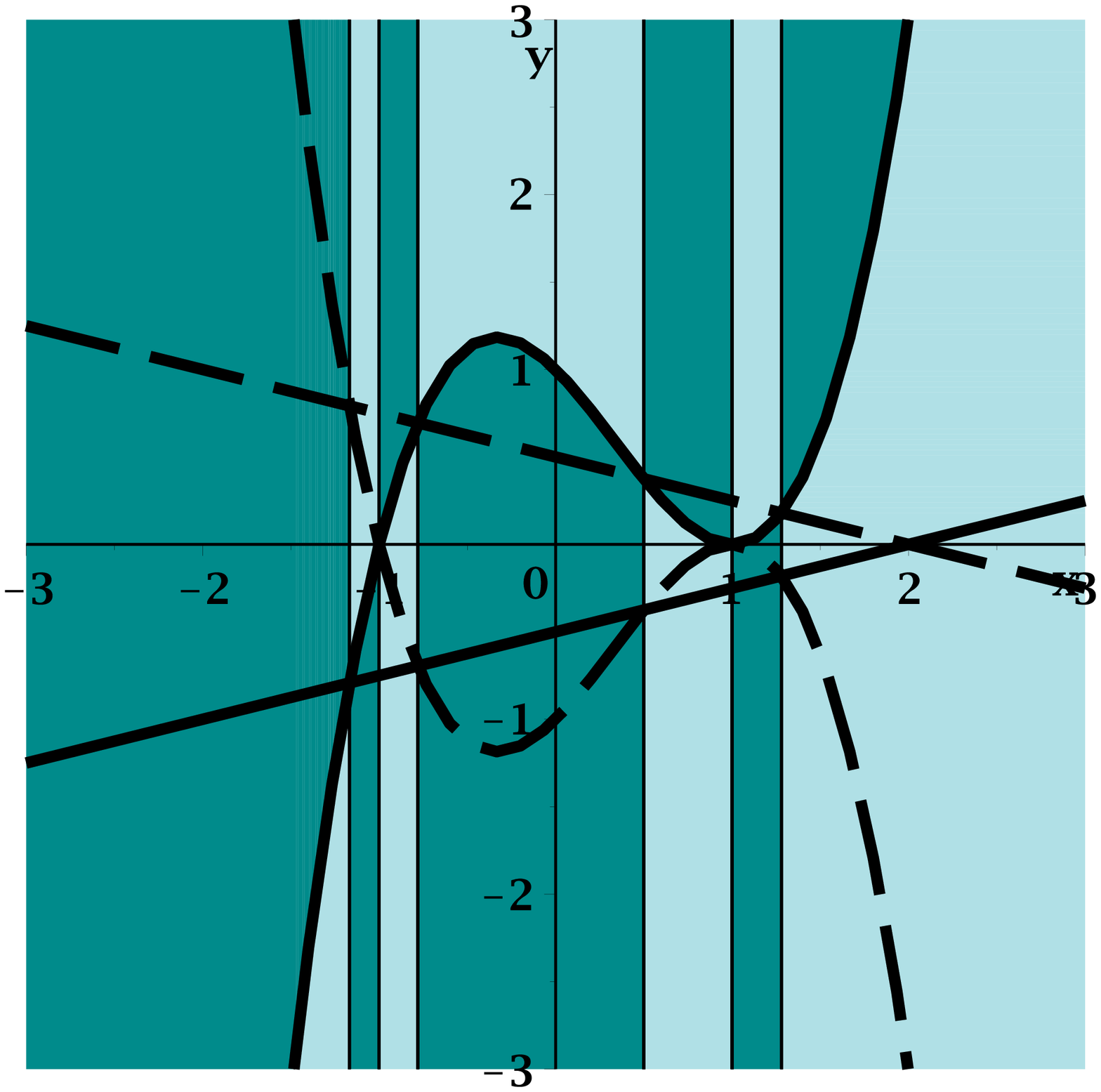}
\includegraphics[width=0.243\textwidth]{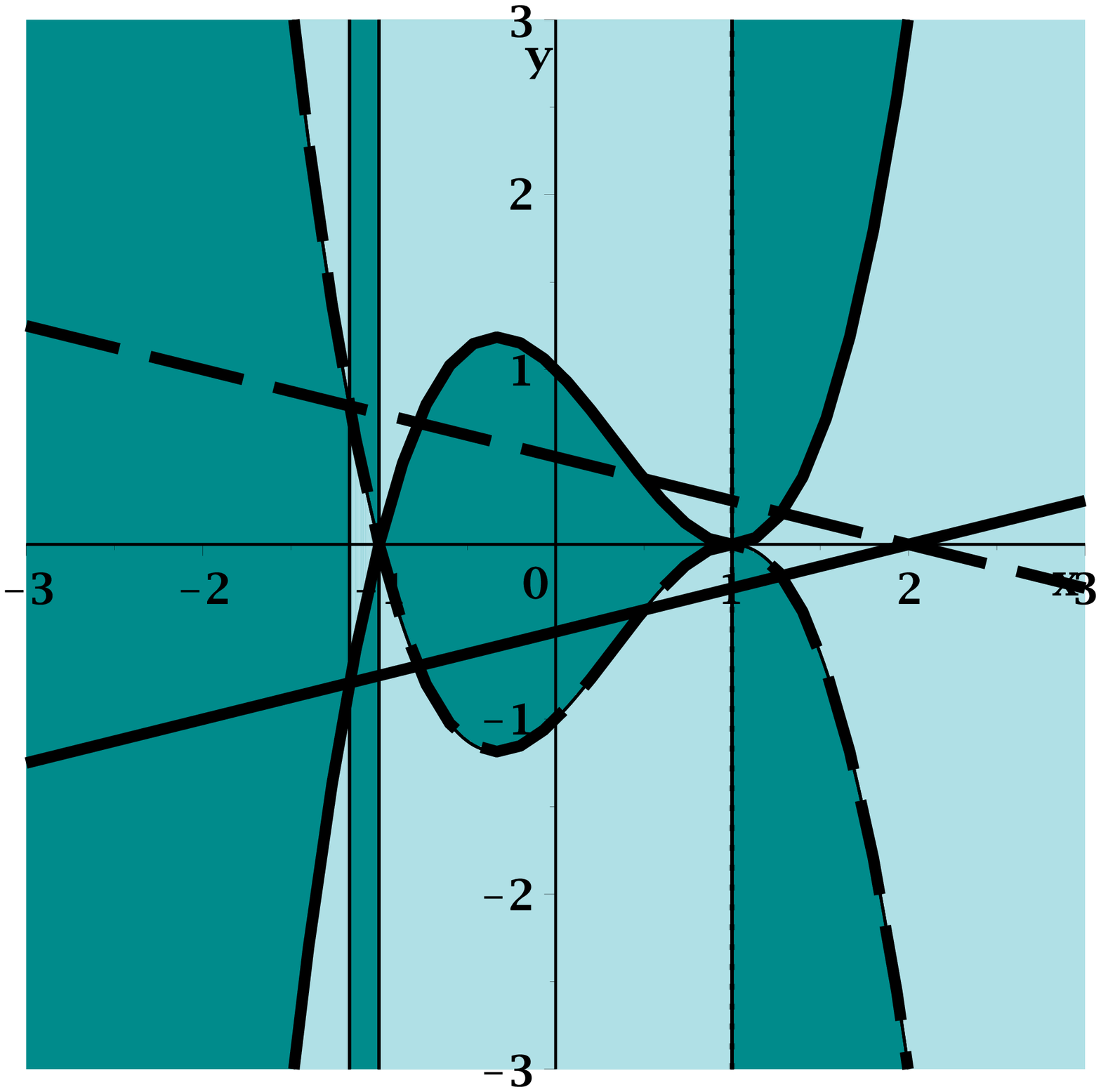}
\includegraphics[width=0.243\textwidth]{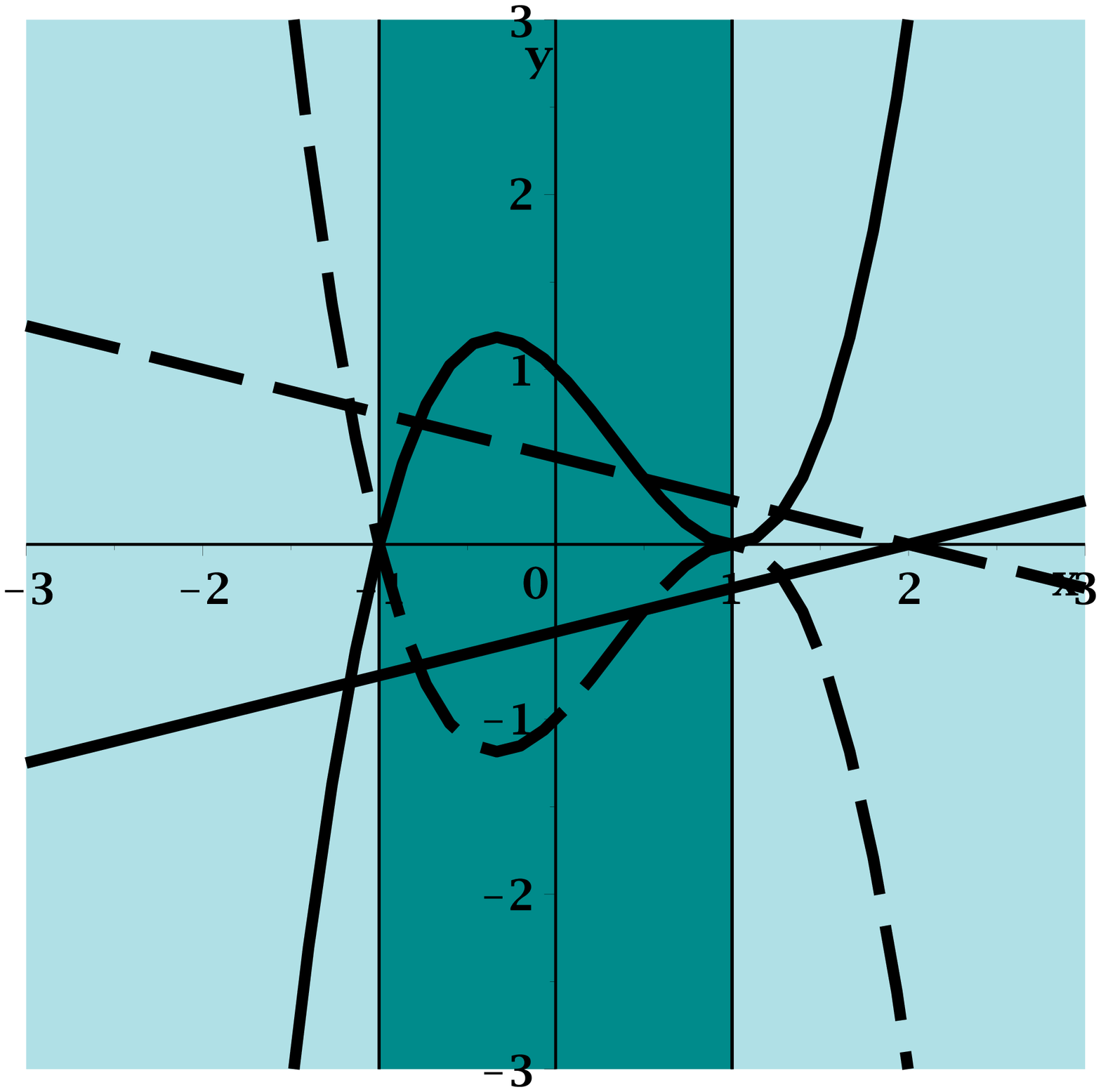}
\includegraphics[width=0.243\textwidth]{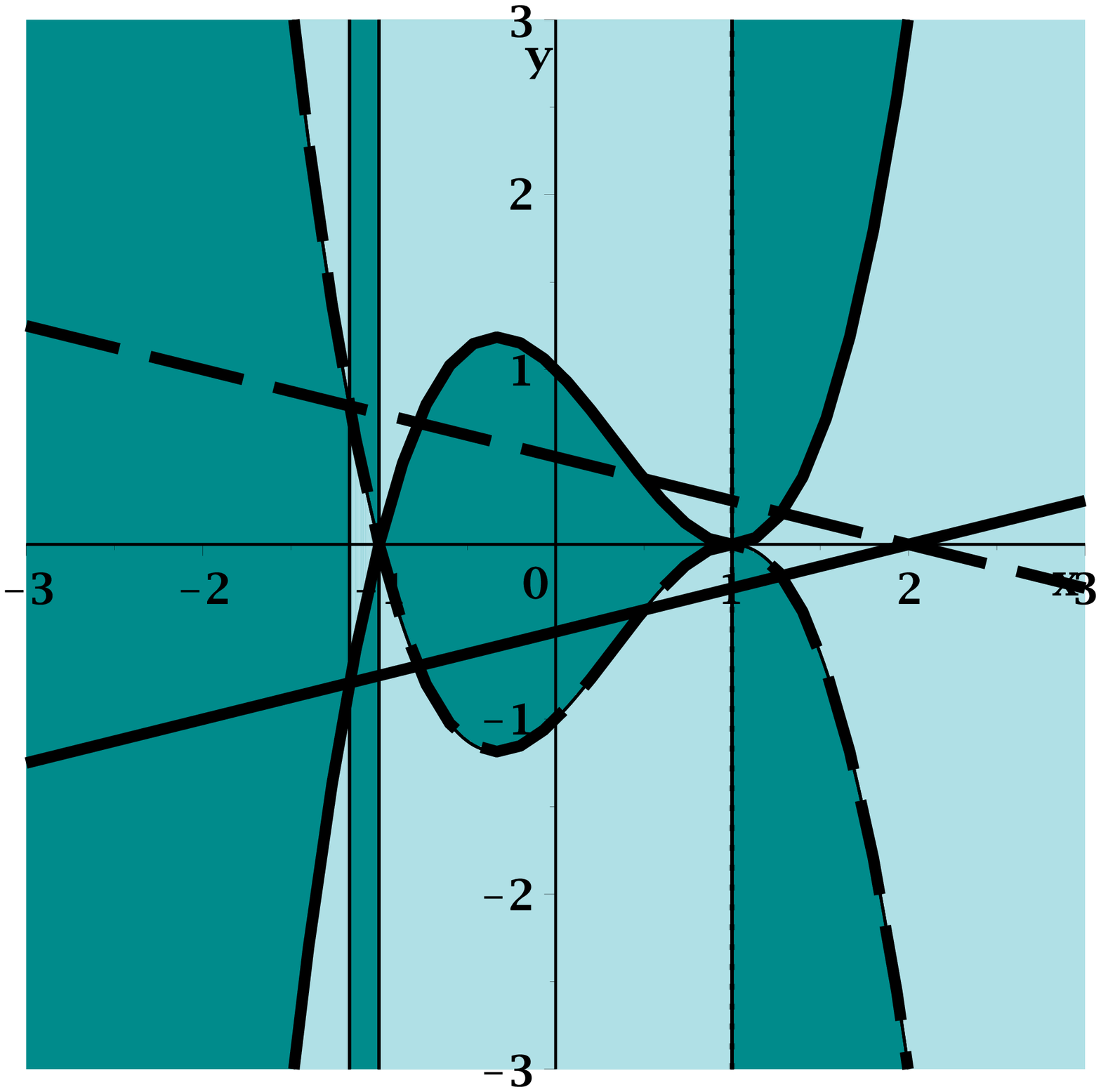}
\vspace*{-15pt}
\end{figure}

\begin{example}[Example 6 in \cite{BDEW13}]
\label{ex:Form}
Let $x \prec y$ and consider the formula $\hat{\Phi}$ above with
\begin{align*}
f_1 &:= (y-1) - x^3+x^2+x, \qquad \quad \psi_1:=g_1<0,  \qquad
g_1 := y - \tfrac{x}{4}+\tfrac{1}{2},  \\
f_2 &:= (-y-1) - x^3+x^2+x, \qquad \, \psi_2:=g_2<0, \qquad
g_2 := -y - \tfrac{x}{4}+\tfrac{1}{2}.
\end{align*} 
The polynomials are plotted in the images of Figure \ref{fig:Form} where the solid curve is $f_1$, the solid line $g_1$, the dashed curve $f_2$ and the dashed line $g_2$.  Various CADs may be computed for this problem: 
\begin{itemize}
\item 
A CAD for $\hat{\Phi}$ using projection and lifting with the operator from \cite{McCallum1999} designating $f_1$:  39 cells as visualised in the first image.
\item As above but designating $f_2$: 39 cells.  Similar to first image but 2 dimensional cell divisions over $f_2$ instead of $f_1$.
\item A TTICAD for $\phi_1, \phi_2$ using projection and lifting with the operator from \cite{BDEMW13}: 31 cells as visualised in the second image.
\item A CAD for $\hat{\Phi}$ using \texttt{RC-TTICAD} (equivalent here to the algorithm in \cite{CM12b}): 9 cells (under any constraint ordering) as visualised in the third image.
\item A TTICAD for $\phi_1, \phi_2$ using \texttt{RC-TTICAD} \cite{BCDEMW14}: 29 cells (under any constraint ordering) as visualised in the fourth image.  
\end{itemize}

The important observation is that $f_1$ has many intersections with $g_2$ and $f_2$ many intersections with $g_1$.  The projection and lifting algorithms can avoid considering those pairs together by splitting into sub-formulae.  In the first image only one EC is sign-invariant while in the second both are, but this was a price worth paying to avoid the intersections.  It is not necessary for \texttt{RC-TTICAD} as this can take advantage of multiple ECs in a formula.  It first identifies the intersection of $f_1$ and $f_2$ and the non-ECs are only considered modulo this set. Hence, even though they are in the same formula, those intersections are never computed.  
\end{example}

In general, \texttt{RC-TTICAD} requires a parent formula to be broken into conjunctive sub-formulae before use, but would never benefit from further decomposition.

\section{Final thoughts}
\label{SEC:Future}

We developed new heuristics to choose constraint orderings for \texttt{RC-TTICAD} \cite{BCDEMW14}, finding that the choice of which to use may depend on the priorities of the user.
A dynamic heuristic (such as one like Heuristic \ref{def:H2} but making choices based on the CCT after each increment) may offer further improvements and will be a topic of future work.

We also revisited other questions of problem formulation questions from \cite{BDEW13}, finding that they did not require further consideration for \texttt{RC-TTICAD}.  However, there was one important issue we did not address, the variable ordering, for which there may be a free or constrained choice.  For example, when using CAD for quantifier elimination we must order the variables as they are quantified but we may change the ordering within quantifier blocks.  It has long been noted that problems which are easy in one variable ordering can be infeasible in another, with \cite{BD07} giving problems where one variable ordering leads to a CAD with a cell count constant in the number of variables and another a cell count doubly exponential.  The analysis was valid for any CAD regardless of the algorithm that produced it and so affects \texttt{RC-TTICAD}.  A key area of our future work will be to analyse how best to choose a variable ordering, and to investigate whether an existing heuristic, or the new ones developed here can help.

\subsection*{Acknowledgements}
This work was supported by the EPSRC (EP/J003247/1), the NSFC (11301524), 
and the CSTC (cstc2013jjys0002).

\end{document}